# Tungsten doping of $Ta_3N_5$-Nanotubes for Band Gap Narrowing and Enhanced Photoelectrochemical Water Splitting Efficiency

Sabina Grigorescu [a], Benedikt Bärhausen [a], Lei Wang [a], Anca Mazare [a], Jeong Eun Yoo [a], Robert Hahn [a], Patrik Schmuki [a,b,*]

[a] *Department of Materials Science and Engineering, WW4-LKO, University of Erlangen-Nürnberg, Martensstrasse 7, D-91058 Erlangen, Germany*

[b] *Department of Chemistry, King Abdulaziz University, Jeddah, Saudi Arabia*

* Corresponding author: e-mail address: schmuki@ww.uni-erlangen.de

**Abstract**

Ordered W-doped $Ta_2O_5$ nanotube arrays were grown by self-organizing electrochemical anodization of TaW alloys with different tungsten concentrations and by a suitable high temperature ammonia treatment, fully converted to W:$Ta_3N_5$ tubular structures. A main effect found is that W doping can decrease the band gap from 2 eV (bare $Ta_3N_5$) down to 1.75 eV. $Ta_3N_5$ nanotubes grown on 0.5 at% W alloy and modified with $(CoOH)_x$ as co-catalyst show ~33% higher photocurrents in photoelectrochemical (PEC) water splitting than pure $Ta_3N_5$.

**Introduction**

Through photoelectrochemical (PEC) water splitting, solar energy can be collected and converted into clean and storable $H_2$ [1-5]. As a photocatalyst, tantalum nitride ($Ta_3N_5$) photoanodes have attracted considerable interest, due to its suitable band gap (~ 2.1 eV), and a theoretical solar-to-hydrogen efficiency as high as 15.9% under AM 1.5G 100 mW cm$^{-2}$ irradiation [4]. However, various carrier recombination pathways lead to significant lower performance in practical devices.

One-dimensional (1D) nanostructures of $Ta_3N_5$ (e.g. nanorods and nanotubes) have the advantages of improving absorbance, promoting the transport and separation of photoexcited charge carriers [5-9]. Recently, photoelectrodes of $Ta_3N_5$ nanotube [10-12] and nanorod [13-16] arrays have been fabricated and showed enhanced PEC activity, demonstrating the high potential of such 1D $Ta_3N_5$ nanostructures.

Furthermore, some successful doping attempts in alkali [5,6] or earth alkali ions [14], have also been shown to further increase mainly the conductivity of $Ta_3N_5$ photoanodes.

In the present work, we use a substrate (alloy based) doping route to in-situ dope the nanotubes during electrochemical growth with a heavy metal ion (tungsten). We demonstrate that W-doping leads to a strong effect on the band-gap of $Ta_3N_5$ and the water splitting activity of self-organized W-doped $Ta_3N_5$ nanotubular photoanodes.

**2. Experimental**

The $Ta_3N_5$ NTs were prepared by electrochemical anodization of metallic Ta foil (0.1 mm thickness, Advent) as well as four different TaW alloys, followed by annealing in ammonia gas. Prior to anodization, TaW alloys were prepared by melting of Ta with different W concentrations (0.5 at%, 1 at%, 2.5 at% and 5 at %) in a vacuum arc furnace. Each alloy was remelted 20 times. The ingots were sliced and polished with a 30 μm diamond disc, cleaned by ultrasonication in acetone and ethanol (5 minutes each), and then dried in a nitrogen stream. Anodization took place in a teflon cell at 60 V for 10 min, using a two electrode configuration. The distance between the working electrode (Ta foil/ TaW alloy) and the counter electrode (Pt foil) was fixed to 1 cm. A mixed solution of 45 ml $H_2SO_4$, 2.5 ml Glycerol, 2.5 ml $H_2O$ and 0.5 g $NH_4F$ was used as the anodization electrolyte. $Ta_3N_5$ nanotubes were obtained by nitridation of $Ta_2O_5$ nanotubes in a horizontal quartz tube furnace under ~200 sccm $NH_3$ flow at 1000 $^0C$ for 2 h, followed by $Co(OH)_x$ as co-catalyst loading as described elsewhere [12].

To evaluate the photoresponse of the nanotubes, photocurrent transients were recorded at a constant potential of 0.4 V (vs. Ag/AgCl) in 1M KOH while the wavelength was varied using a Xe lamp and an Oriel Cornestone 130 1/8 m motor driven monochromator.

Morphological characterization (SEM), compositional analysis (XPS,EDX), structural information (XRD) as well as PEC water-splitting characterization were carried out as described in previous work [12].

**3. Results and Discussions**

Figure 1 a-c upper part shows the SEM cross-section and the top view as inset of the as-formed $Ta_2O_5$ nanotube layers on pure tantalum, 0.5 at% W and 5 at% W alloy. Ordered W-doped $Ta_2O_5$ nanotubes are formed on the alloy surface having an average length of 5.5 μm, with length variations of the nanotubes between 3.5 μm and 7.7 μm; the top layer shows a porous initiation layer typical for anodization. Under the used anodization conditions, the nanotube layer is very stable on the metallic substrate and no difference in the morphology of nanotube surfaces was observed for the anodized alloys and pure tantalum.

After nitridation, as shown in Figure 1a-c lower part, the overall tubular structure was maintained but a change in the nanotube wall morphology is observed, that is, the average layer thickness decreases from 5.5 μm to 4.2 μm due to a higher molar density of Ta in $Ta_3N_5$ than in $Ta_2O_5$ [8].

For all samples (Ta or TaW alloys) it was found that the as-grown nanotube layers are amorphous, but are converted to a crystalline structure by high temperature nitridation. Figure 1d shows examples of the XRD patterns for undoped, 0.5 at% W and 5 at% W alloy $Ta_3N_5$ nanotubes. The main identified phase of $Ta_3N_5$ NTs is the orthorhombic phase, but also some other intermediate phases summarized and labeled as $Ta_xN_y$ (TaN, $Ta_2N$, etc.) are present. For all nanotubes obtained on different W-content alloys similar diffractograms are registered. No peaks related to tungsten nitrides (e.g. $W_2N$ or WN) could be identified. In Fig. 1e a more detailed evaluation of the XRD data (as an example the (110) peak at 24,52° is shown) is given by plotting the deviation of the peak position and the full width half maximum (FWHM) as a function of the W concentration. With increasing tungsten concentration a shift in the peak position towards smaller diffraction angles accompanied by a broadening of the FWHM is observed. The plateaus in the slopes of the XRD results suggest that just a limited amount of W-atoms can be build into the crystal lattice.

In order to study the chemical composition of the W-doped $Ta_3N_5$ nanotubes, XPS and EDX measurements were performed. As shown in Figure 2, EDX and XPS reveal a linear increase of the W concentration in the tubes with the W-concentration in the alloy. It is evident that the final tungsten concentration in the nanotubes is slightly smaller than in the corresponding alloy, as some of the tungsten species seem to be etched out and lost into the electrolyte during anodic growth. Nevertheless, the results show that by using different W-containing alloys for anodization, a desired doping concentration in the nanotubes can be established.

In order to assess the effect of W-doping on the electronic and optical properties of the materials, photocurrent spectra were recorded for undoped and W-doped $Ta_3N_5$ nanotubes.

Figure 2b depicts the evaluation of the bandgaps of W-doped and undoped nanotubes using photocurrent measurements. Clearly, the photo-electrochemical measurements confirm that tungsten doping of tantalum nitride leads to a lowering in the band gap from 2eV down to 1.75 eV for all W concentrations. This change in optical properties can be also observed by eye (see optical images shown as insets in the figure), where the color changes from red (for undoped samples) over purple (0.5 at% W) to grey (5 at%). Figure 2c shows incident photon to current efficiency (IPCE) spectra of the nanotubular photoanodes with different W concentrations. Despite the revealed IPCE response at higher wavelengths of the tantalum nitride nanotubes doped with small amounts of tungsten, these photoanodes show, in comparison to undoped material, also a better performance at lower wavelengths. On the other hand, doping with higher amounts of tungsten leads to a decrease in the IPCE values. This decrease may be attributed to the higher amount of defect states created by the doping species, which may increase the recombination rate of charge carriers.

Photoelectrochemical water splitting experiments were performed under simulated sunlight AM 1.5 at 100 mW cm$^{-2}$ conditions. Figure 3a shows the photocurrent transients of the $Ta_3N_5$ nanotube layers for undoped tantalum, 0.5 at% W and 5 at% W alloy with $Co(OH)_x$ as co-catalyst in 1 M KOH (pH 13.7), while Figure 3b shows the comparison for the photocurrent density at 400 mV (vs. Ag/AgCl) between pure $Ta_3N_5$ and all tested alloys with and without co-catalyst.

In comparison to the pure $Ta_3N_5$, an enhancement in the photoelectrochemical current response of about 0.85 mAcm$^{-2}$ or ~33% is observed for the 0.5 at% W alloy with $Co(OH)_x$ as co-catalyst (Figure 3b). The 0.5 at% W alloy shows drastically higher efficiency when compared to the co-catalyst free W-doped sample. Moreover, the high rate of recombination observed in the range between -0.2 and 0.4 V (vs.Ag/AgCl) by chopping the light (shown in Figure 3a) reveals the high potential of this material, but further improvement to suppress recombination is needed.

For all W-doped $Ta_3N_5$ nanotubes, the photocurrent response is higher with the $Co(OH)_x$ modification of the nanotube surface (see Figure 3b). However, the water splitting efficiency decreases with the increase of W-concentration; this can be attributed to the higher amount of defect states created by the doping amount, which strongly influences the recombination rate, in line with IPCE measurements.

## 4. Summary

In the current work we show a successful band-gap reduction in $Ta_3N_5$ nanotubes by W doping. The doping process occurs in-situ (i.e. during the anodic growth of a Ta-W- oxide from a TaW substrate). It should be noted that many parameters of the process (nanotube morphology, conversion conditions to nitride such as temperature and time, co-catalyst loading etc.) need to be further optimized in order to explore the full potential of this process. However, from the present results it is clear that $Ta_3N_5$ nanotubes doped with a small amount of tungsten ions can significantly enhance the photoelectrochemical water splitting performance of Ta-nitride material.

**Acknowledgments**

The authors would like to acknowledge financial support from ERC, DFG within the SPP 1613 and the Cluster of Excellence (Engineering of Advanced Materials) of the FAU.

**References:**

[1]     A. Fujishima, K. Honda, Electrochemical photolysis of water at a semiconductor electrode, Nature 238 (1972) 37–38.

[2]     M. Grätzel, Solar water splitting cells, Nature 414 (2001) 338-344.

[3]     M. G. Walter, E. L. Warren, J. R. McKone, S. W. Boettcher, Q. X. Mi, E. A. Santori, N. S. Lewis, Solar Water Splitting Cells, Chemical Reviews 110 (2010) 6446-6473.

[4]     M. Tabata, K. Maeda, M. Higashi, D. Lu, T. Takata, R. Abe, K. Domen, Modified $Ta_3N_5$ powder as a photocatalyst for $O_2$ evolution in a two-step water splitting system with an iodate/iodide shuttle redox mediator under visible light, Langmuir 26 (2010) 9161-9165.

[5]     Y. Kado, R. Hahn, C.-Y. Lee, P. Schmuki, Strongly enhanced photocatalytic water splitting efficiency for Na doped $Ta_3N_5$ − nano channel structure, Electrochem.Commun., 17 (2012) 67-70.

[6]     Y. Kado, C.-Y. Lee, K. Lee, J. Müller, M. Moll, E. Spiecker, P. Schmuki, Enhanced water splitting activity of M-doped $Ta_3N_5$ (M = Na, K, Rb, Cs), Chem. Commun. 48 (2012) 8685-8687.

[7]     Z. Su, L. Wang, S. Grigorescu, K. Lee, P. Schmuki, Hydrothermal growth of highly oriented single crystalline $Ta_2O_5$ nanorod arrays and their conversion to $Ta_3N_5$ for efficient solar driven water splitting, Chem. Commun. (2014) accepted.

[8]     M. Li, W. Luo, D. Cao, X. Zhao, Z. Li, T. Yu, Z. Zou, A co-catalyst-loaded $Ta_3N_5$ photoanode with a high solar photocurrent for water splitting upon facile removal of the surface layer, Angew. Chem. Int. Ed. 52 (2013) 11016–11020.

[9]     E. Nurlaela, S. Ould-Chikh, M. Harb, S. del Gobbo, M. Aouine, E. Puzenat, P. Sautet, K. Domen, J.-M. Basset, K. Takanabe, Critical role of the semiconductor−electrolyte interface

in photocatalytic performance for water-splitting reactions using $Ta_3N_5$ particles, Chem. Mater. 26 (2014) 4812−4825.

[10]    X. Feng, T.J. LaTempa, J.I. Bastian, G.K. Mor, O.K. Varghese, C.A. Grimes, $Ta_3N_5$ nanotube arrays for visible light water photoelectrolysis, Nano Lett. 10 (2010) 948–952.

[11]    Y. Cong, H.S. Park, S. Wang, H.X. Dang, F.-R. F. Fan, C.B. Mullins, A.J. Bard, Synthesis of $Ta_3N_5$ nanotube arrays modified with electrocatalysts for photoelectrochemical water oxidation, J. Phys. Chem. C 116 (2012) 14541–14550.

[12]    Z. Su, S. Grigorescu, L. Wang, K. Lee, P. Schmuki, Fast fabrication of $Ta_2O_5$ nanotube arrays and their conversion to $Ta_3N_5$ for efficient solar driven water splitting, Electrochem. Commun. (2014) accepted.

[13]    Y. Li, T. Takata, D. Cha, K. Takanabe, T. Minegishi, J. Kubota, K. Domen, Vertically aligned $Ta_3N_5$ nanorod arrays for solar-driven photoelectrochemical water splitting, Adv. Mater. 25 (2013) 125–131.

[14]    Y. Li, L. Zhang, A. Torres-Pardo, J.M. González-Calbet, Y. Ma, P. Oleynikov, O. Terasaki, S. Asahina, M. Shima, D. Cha, L. Zhao, K. Takanabe, J. Kubota, K. Domen, Cobalt phosphate-modified barium-doped tantalum nitride nanorod photoanode with 1.5% solar energy conversion efficiency, Nat. Comm. 4 (2013) 2566.

[15]    C. Zhen, L.Z. Wang, G. Liu, G.Q. Lu (Max), H.M. Cheng, Template-free synthesis of $Ta_3N_5$ nanorod arrays for efficient photoelectrochemical water splitting, Chem. Commun. 49 (2013) 3019–3021.

[16]    J. Hou, Z. Wang, C. Yang, H. Cheng, S. Jiao, H. Zhu, Cobalt-bilayer catalyst decorated $Ta_3N_5$ nanorod arrays as integrated electrodes for photoelectrochemical water oxidation, Energy Environ. Sci. 6 (2013) 3322–3330.

**Figure captions:**

Figure 1. Upper images are cross sections and top (inset) SEM images of (a) pure tantalum, (b) 0.5 at% W and (c) 5 at% W alloy $Ta_2O_5$ nanotubes, and lower part shows the corresponding $Ta_3N_5$ nanotubes; (d) XRD patterns for the pure tantalum, 0.5 at% W and 5 at% W alloy $Ta_3N_5$ nanotubes; (e) detailed evaluation of the XRD data using the (110) peak.

Figure 2. (a) Linear dependence of the W concentrations added to the alloys and the resulting atomic concentration of tungsten doped into tantalum nitride nanotubes measured by EDX and XPS; (b) photo-electrochemical measurements of co-catalyst loaded undoped and W-doped $Ta_3N_5$ nanotube (different tungsten concentration, insets are optical images) and (c) photocurrent spectra plotted as IPCE vs. wavelength of the co-catalyst loaded undoped and doped $Ta_3N_5$ nanotube layers. Photoelectrochemical experiments were conducted in 1M KOH at an applied potential of 400mV (vs. Ag/AgCl).

Figure 3. (a) Current-potential curve of the pure tantalum, 0.5 at% W and 5 at% W alloy alloy $Ta_3N_5$ nanotube layers with $Co(OH)_x$ as co-catalyst under chopped AM 1.5G simulated sunlight in 1 M KOH solution (pH=13.7) and a scan rate of 2 mV s$^{-1}$; (b) photocurrent densities at 400 mV (vs. Ag/AgCl) for pure and W-doped $Ta_3N_5$ nanotube photoanodes with and without co-catalyst.

**Figure 1**

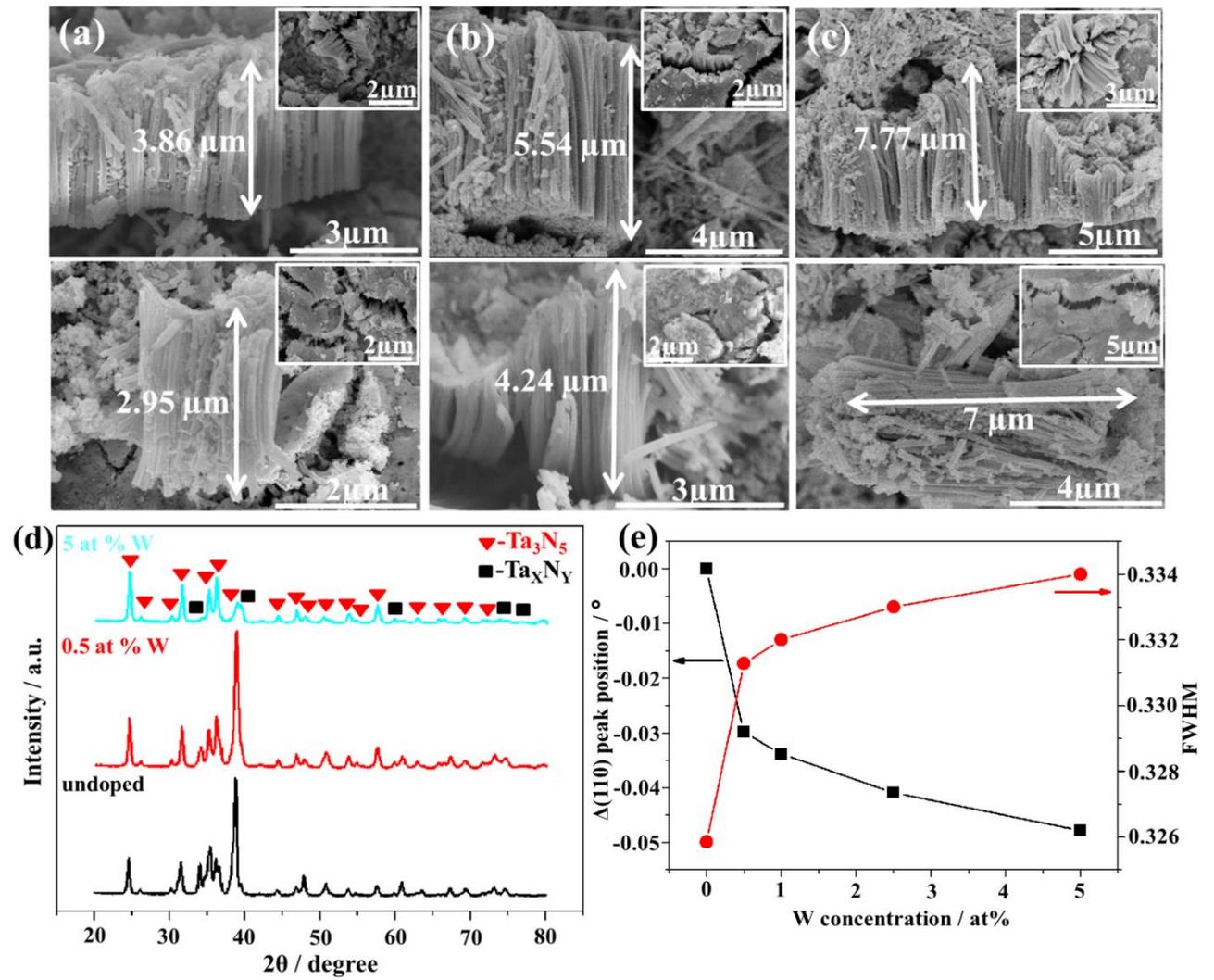

**Figure 2**

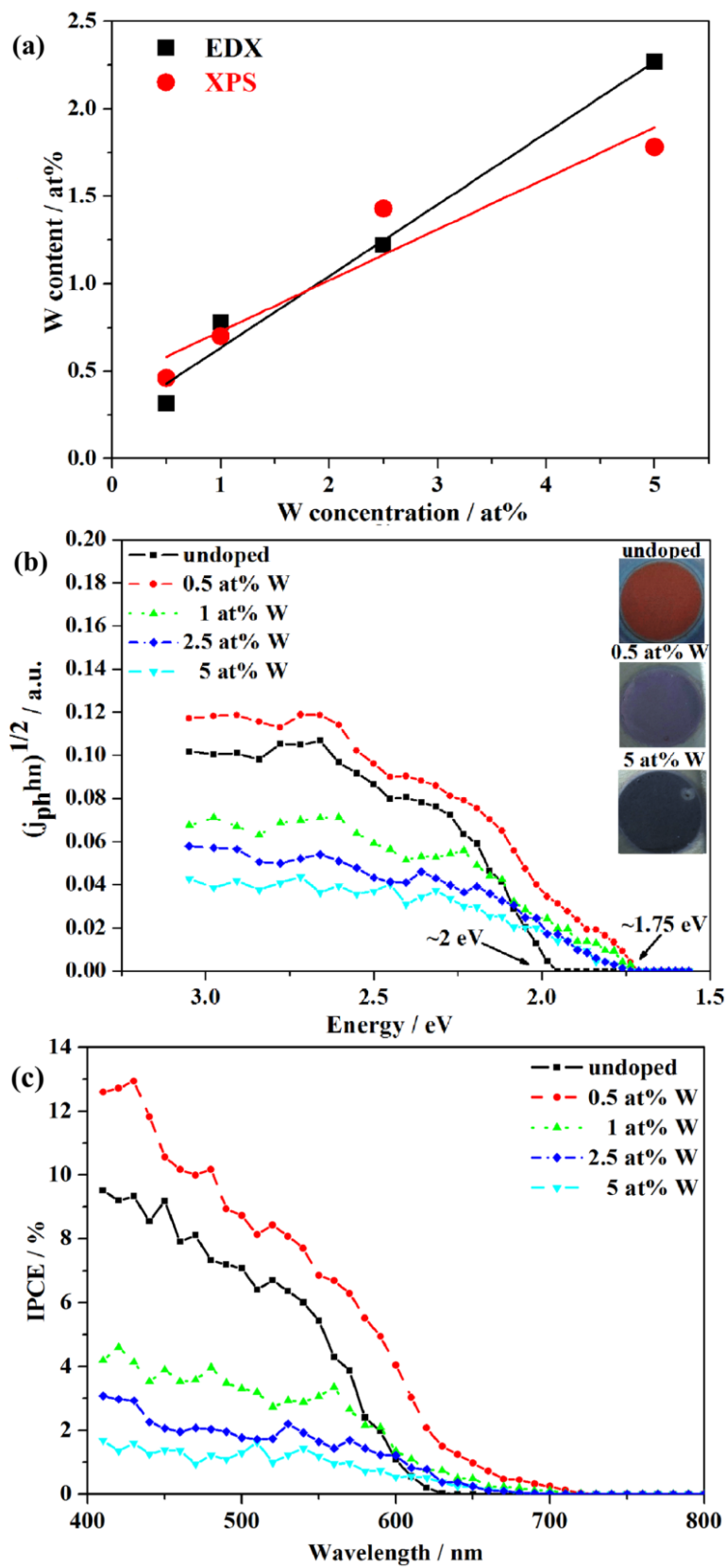

**Figure 3**

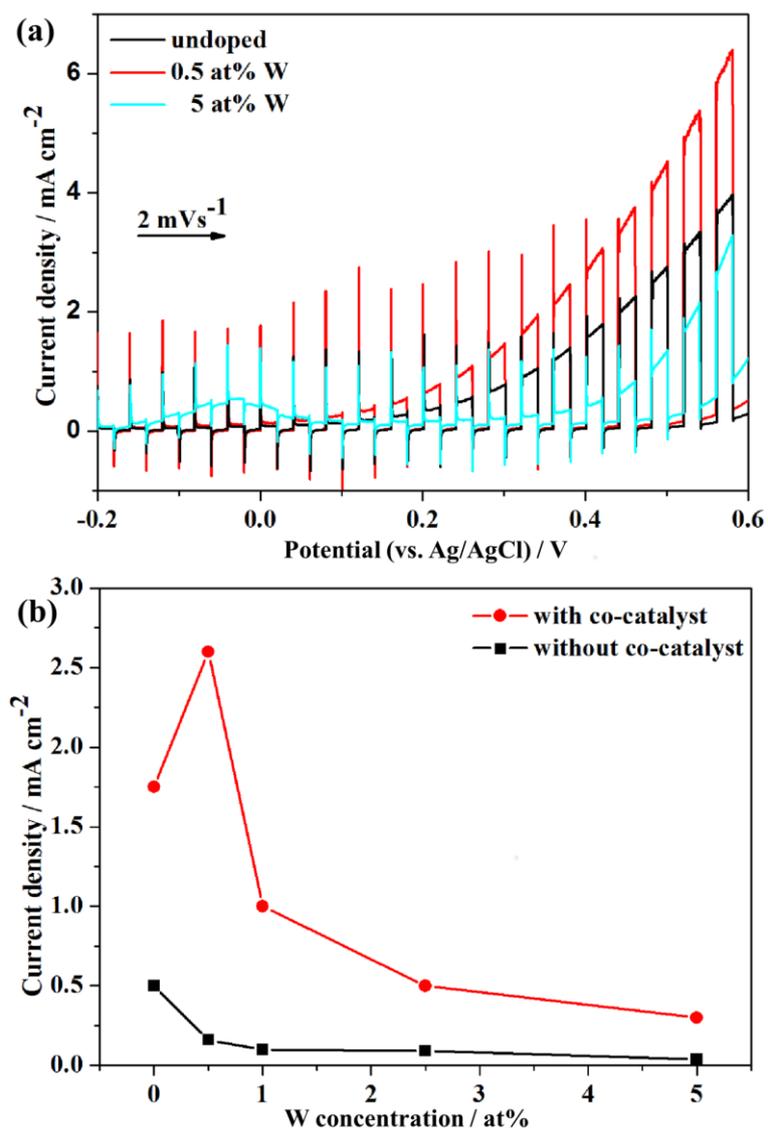